\documentclass{emulateapj}

\shorttitle{The \ion{He}{2} Break}
\shortauthors{Syphers et al.}

\begin{document}

\title{{\em HST}\altaffilmark{1} Spectral Observations near the \ion{He}{2} Ly$\alpha$ Break: Implications for \ion{He}{2} Reionization}

\author{David Syphers\altaffilmark{2,3},
Scott F. Anderson\altaffilmark{4},
Wei Zheng\altaffilmark{5},
Avery Meiksin\altaffilmark{6},
Daryl Haggard\altaffilmark{4},
Donald P. Schneider\altaffilmark{7},
Donald G. York\altaffilmark{8,9}
}

\altaffiltext{1}{Based on observations with the NASA/ESA \textit{Hubble Space Telescope} obtained at the Space Telescope Science Institute, which is operated by the Association of Universities for Research in Astronomy, Inc., under NASA contract NAS5-26555.}

\altaffiltext{2}{CASA, Department of Astrophysical and Planetary Sciences, University of Colorado, Boulder, CO 80309, USA; David.Syphers@colorado.edu}

\altaffiltext{3}{Physics Department, University of Washington, Seattle, WA 98195, USA}

\altaffiltext{4}{Astronomy Department, University of Washington, Seattle, WA 98195, USA; anderson@astro.washington.edu}

\altaffiltext{5}{Department of Physics and Astronomy, Johns Hopkins University, Baltimore, MD 21218, USA; zheng@pha.jhu.edu}

\altaffiltext{6}{Scottish Universities Physics Alliance (SUPA), Institute for Astronomy, University of Edinburgh, Royal Observatory, Edinburgh EH9 3HJ, UK}

\altaffiltext{7}{Department of Physics \& Astronomy, Pennsylvania State University, 525 Davey Lab, University Park, PA 16802, USA}

\altaffiltext{8}{Department of Astronomy and Astrophysics, The University of Chicago, 5640 South Ellis Avenue, Chicago, IL 60637, USA}

\altaffiltext{9}{Enrico Fermi Institute, The University of Chicago, 5640 South Ellis Avenue, Chicago, IL 60637, USA}

\begin{abstract}

Quasars that allow the study of IGM \ion{He}{2} are very rare, since they must be at high redshift along sightlines free of substantial hydrogen absorption, but recent work has dramatically expanded the number of such quasars known.
We analyze two dozen higher-redshift ($z=3.1$--$3.9$) low-resolution \ion{He}{2} quasar spectra from \textit{HST} and find that their \ion{He}{2} Gunn-Peterson troughs suggest exclusion of very early and very late reionization models, favoring a reionization redshift of $z \sim 3$.
Although the data quality is not sufficient to reveal details such as the expected redshift evolution of helium opacity, we obtain the first ensemble measure of helium opacity at high redshift averaged over many sightlines: $\tau=4.90$ at $z\sim3.3$.
We also find that it would be very difficult to observe the IGM red wing of absorption from the beginning of \ion{He}{2} reionization, but depending on the redshift of reionization and the size of ionization zones, it might be possible to do so in some objects with the current generation of UV spectrographs.

\end{abstract}


\section{Introduction}
\label{sec:intro}

Reionization of the intergalactic medium (IGM) is a major cosmological milestone in the evolution of the universe.
Although reionization of hydrogen and \ion{He}{1} occurred at $z>6$ \citep[e.g.,][]{fan06,kom10}, reionization of \ion{He}{2} was delayed.
Stars, the initial source of much of the ionizing radiation, do not produce enough high-energy UV photons to ionize \ion{He}{2}, and thus this process occurred only when there was substantial hard ionizing radiation from quasars.
Both theoretically and observationally, \ion{He}{2} reionization is thought to have occurred gradually at redshifts $\sim$3--4 \citep[e.g.,][]{sok02,aga05}.
Even under the hard photoionizing conditions at $z\sim 2$--$4$, \ion{He}{2} outnumbers \ion{H}{1} by a factor of $\eta \sim 50$--100 \citep{zhe04b,fec06}, and thus \ion{He}{2} has much stronger opacity than hydrogen.
As a result, the most sensitive and direct probe of this reionization era is the \ion{He}{2} Gunn-Peterson effect, analogous to the effect in hydrogen where a substantially neutral IGM causes Ly$\alpha$ line blanketing to create a trough in quasar spectra \citep{GP65}.

There is a variety of indirect evidence that \ion{He}{2} reionization occurred near $z \sim 3$.
Much of this work has tried to exploit the IGM heating associated with this epoch, and its effect on the \ion{H}{1} Ly$\alpha$ forest, either by examining the increased temperature using various statistics related to thermal broadening \citep[e.g.,][]{sch00,lid09,bec10}, or the opacity changes associated with heating \citep[e.g.,][]{ber03,fau08}.
While suggestive of reionization at $z \sim 3$, such studies are fraught with difficult systematics, contradicting observations \citep[e.g.,][]{mcd01,kim02}, and difficult theoretical interpretations \citep{bol09,mcq09a}.
Another approach has been to use ion ratios to determine the ionizing UV background on either side of the \ion{He}{2} Lyman limit (228~\AA), which should change as the helium opacity changes \citep[although see][]{fur09}.
Such measurements are also difficult and suffer from substantial uncertainty, but they generally agree with a $z \sim 3$ reionization \citep[e.g.,][]{son98,aga07}.

Because of the large uncertainty associated with these indirect measurements, it is valuable to pursue direct measurements of the \ion{He}{2} opacity, as an independent method of constraining the redshift of reionization.
Unlike the hydrogen Gunn-Peterson trough, which saturates at $x_{\textrm{{\scriptsize H}}\,\textrm{{\scriptsize{\sc I}}}} \sim 10^{-5}$ (\citealp{fan06}, although see \citealp{mes10}), the \ion{He}{2} Ly$\alpha$ Gunn-Peterson trough can remain sensitive to ion fractions $x_{\textrm{{\scriptsize He}}\,\textrm{{\scriptsize{\sc II}}}} \gtrsim 0.01$, due to the later redshift of its reionization, the lower abundance of helium versus hydrogen, the shorter wavelength of the line ($\tau_{{\rm GP}} \sim \lambda_{{\rm Ly}\alpha}$), and density fluctuations in the IGM \citep{mcq09}.
In addition, the large fluctuations in the ionizing background from the rare nature of the luminous quasars thought to be responsible for \ion{He}{2} reionization make transmission possible even during the early stages of reionization \citep{fur09}.

With \ion{He}{2}~Ly$\alpha$ at $303.78$~\AA, observations of the Gunn-Peterson trough at the relevant redshifts must take place in the far UV, and are thus very sensitive to low-redshift hydrogen absorption.
Since observations in this regime must take place in space, much of this work has been done with the {\it Hubble Space Telescope} ({\it HST}), although observations of a few brighter targets were possible with the {\it Hopkins Ultraviolet Telescope} ({\it HUT}) and the {\it Far Ultraviolet Spectroscopic Explorer} ({\it FUSE}).
The recently installed Cosmic Origins Spectrograph (COS) on {\it HST} is proving especially well suited for this work.
Due to intervening IGM hydrogen, only a few percent of randomly selected $z \sim 3$ quasar sightlines are clean (no substantial \ion{H}{1} opacity) down to the \ion{He}{2} break \citep{mol90}.
Verified \ion{He}{2} quasars are therefore rare, although the advent of the \textit{Galaxy Evolution Explorer} ({\it GALEX}) UV maps covering most of the sky has recently greatly increased selection efficiency \citep[][hereafter S09a,b]{syp09a,syp09b}.
The difficulty of probing the redshift regime interesting for \ion{He}{2} reionization is increased because the number density of luminous quasars may decline exponentially in $z=3$--$4.5$ \citep{osm82,sch95,ric06,bru09}.

To date there are 27 quasars observed to be free of substantial absorption down to the \ion{He}{2} Ly$\alpha$ break (with a few others that have observable flux at the break but also substantial intervening absorption).
Six of these have relatively high S/N spectral observations: Q0302-003 \citep[$z=3.29$,][]{hog97,hea00}, HE~2347-4342 \citep[$z=2.90$,][]{sme02,zhe04b,shu10}, HS~1700+6416 \citep[$z=2.72$,][]{fec06}, PKS1935-692 \citep[$z=3.18$,][]{and99}, SDSS~J2346-0016 ($z=3.5$), and SDSS~J1711+6052 ($z=3.8$) \citep[both][hereafter Z08]{zhe08}.
Our team recently found 19 others, ranging from $z \sim 3.1$--$3.8$, with reconnaissance using the {\it HST} Advanced Camera for Surveys/Solar Blind Channel (ACS/SBC) prism (S09a,b).
The remaining two known \ion{He}{2} quasars are HS~1157+3143 \citep[$z=3.00$,][]{rei05} and SDSS~J1614+4859 \citep[$z=3.8$,][]{zhe05}, the latter of which was reobserved in S09a at increased (though still not high) S/N.

In section \ref{sec:stacks_wing}, we use this growing number of \ion{He}{2} quasars to make spectral stacks in redshift bins, thereby averaging over sightline and individual object variations.
This averaging is a vital step, since the sightline variation is expected to be substantial during \ion{He}{2} reionization due to the fact that the ionizing photons come from rare, luminous quasars \citep{fur09} and the IGM is inhomogeneous \citep{mcq09}.
Using these stacks, we can start to answer whether or not seemingly unusual absorption profiles, such as that of SDSS~J1711+6052 (Z08), are indicative of overall IGM evolution, or simply happenstance intervening absorption along particular sightlines.
We discuss the feasibility of observing IGM red wing absorption, and comment briefly on apparent emission lines in our spectra.

In section \ref{sec:igm_opacity}, we use our substantially expanded sample of \ion{He}{2} quasars from S09a,b to determine the IGM helium opacity.
We consider fits to the spectral stacks of section \ref{sec:stacks_wing}, as well as aggregate statistics using pixel-by-pixel opacity measurements of individual objects.
Understanding the redshift evolution of \ion{He}{2} sightlines would shed light on the progress of \ion{He}{2} reionization, so we here begin this consideration.
We find that while our current data can provide the first good estimate of the helium Gunn-Peterson opacity averaged over a broad high-redshift range, they give no conclusive evidence about any redshift evolution.
The methodology we establish will be applicable to future higher S/N observations.
We conclude in section \ref{sec:conclusion}.

\section{Averaged Quasar Spectra and the Proximity Profile}
\label{sec:stacks_wing}

We conducted surveys in {\it HST} cycles 15, 16, and 16 supplemental (GO programs 10907, 11215, and 11982) searching for \ion{He}{2} quasars with the ACS/SBC prism PR130L.
This prism covers roughly 1250~\AA\ ($R=380$) to 1850~\AA\ ($R=40$).
The spectra were extracted using aXe 1.6 \citep{kum09}; further details may be found in S09a,b.
The targeted quasars were identified in the Sloan Digital Sky Survey \citep[SDSS;][]{yor00}, and detected as UV sources in the {\it GALEX} surveys \citep[the all-sky AIS, the medium-depth MIS, or the deep DIS;][]{mor07}.
The cross-correlation catalog thus created is presented and described in S09a.
In this section we consider methods of combining these spectra to extract information about typical \ion{He}{2} quasars.

\subsection{Stacks of Spectra}
\label{sec:spectrum_stacks}

Our $\sim$4 ks {\it HST} exposures were intended as reconnaissance to verify flux all the way down to \ion{He}{2}~Ly$\alpha$, but stacked spectra and a few higher flux quasars allow more detailed analysis, which we pursue here.
The individual spectra for all objects can be found in S09a,b, along with details on the observations and data reduction.
We found 29 UV-bright quasars at $z>3.1$, at least 23 of which have flux breaks at \ion{He}{2} Ly$\alpha$, and 19 of which are free of any observed absorption redward of the \ion{He}{2} break.
Of those whose flux breaks redward of \ion{He}{2}~Ly$\alpha$, one has a clear continuum break to zero detectable flux $\sim$15~\AA\ (rest) redward of the expected 304~\AA\ break, likely due to a low-redshift hydrogen Lyman-limit system (LLS), while the others have more extended, complicated absorption, perhaps arising from hydrogen absorption line blanketing near the Lyman limit of an LLS or a damped Ly$\alpha$ absorber (DLA).
Two of the 23 with breaks have relatively strong emission lines, with weak, low-S/N continua.
Z08 contains two other high-S/N ACS prism spectra that we use in our analysis.

In order to average over sightline variance and extract information from the aggregate spectra, we coadd the spectra in redshift bins.
We performed an observed-frame stack of these spectra in S09b, to verify that ACS/SBC PR130L does not have any strong instrumental features uncorrected in the reduction.
In S09a, we presented median rest-frame stacks of the 31 far-UV-bright spectra in three large redshift bins ($3.1 < z < 3.3$, $3.3 \leq z < 3.6$, and $3.6 \leq z < 4.1$), finding no obvious redshift evolution, but pursuing no detailed analysis.

In this paper, we improve the coadding method, and present a more detailed analysis of the resultant stacks.
The resolution of ACS spectra depends in a highly nonlinear manner on wavelength.
Because each quasar is at a distinct redshift, the restframe wavelength associated with each pixel (the knots of a spline fit), as well as the restframe wavelength range spanned by each such pixel (i.e., the dispersion solutions transformed to the restframes), will differ for each quasar to be combined in the spectral stack.
We must, therefore, map the spectra to a common wavelength grid before coadding.
In S09a, we did this by spline fitting spectra to the wavelength knots of the highest redshift (lowest resolution) quasar.
This is not a bad approximation, but is suboptimal for a number of reasons.
It limits the upper wavelength in the rest frame to that of the highest-redshift object, ignoring data we have beyond that; it neglects any finer detail that we can extract from the higher-resolution spectra, since it uses the lowest-resolution wavelength knots; in some circumstances, there is not an obvious way to extract an error spectrum for the resultant stack; and finally, while the spline-fitting method is good for wavelength knots near each other, it is somewhat worse for the substantially shifted knots we have here.

Our approach in the current work is to make the spectra almost everywhere continuous by replacing each discrete spectrum by the sum of a set of top-hat functions, one for each pixel.
We then define a wavelength grid, and for each pixel of this grid, calculate the overlap with every pixel in the individual spectra (adding all contributions in each individual spectrum, weighted by their overlaps).
We then take the median over all the input spectra for each pixel in the stacked spectrum.
We considered other methods, such as taking the mean or a S/N-weighed mean instead of the median, but we found in simulations of noisy spectra that the median is marginally better even when there is no sightline variance.
With sightline variance, the median is certainly more robust.

\begin{figure}
\epsscale{1.1}
\plotone{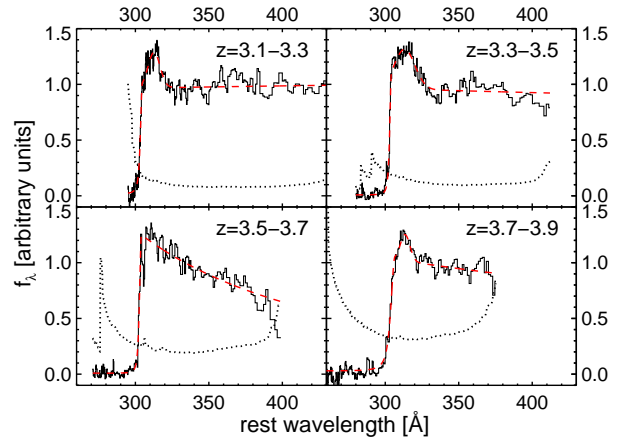}
\caption{Stacked spectra in redshifts bins with fits overplotted (dashed red lines and spectrum errors shown (dotted lines).
There are 7 objects in redshift $3.1$--$3.3$, 7 objects in redshift $3.3$--$3.5$, 5 objects in redshift $3.5$--$3.7$, and 6 objects in redshift $3.7$--$3.9$.}
\label{fig:stacked_specfits}
\end{figure}

Prior to this coadding, we shift the spectra to their respective rest frames, using the SDSS pipeline redshifts (verified by eye to be accurate, except in one case where we shifted an object by $\Delta z = 0.05$).
Then we normalize each spectrum by its continuum flux (avoiding any broad rise towards \ion{He}{2}~Ly$\alpha$).
In the rest frame, we stack the 25 spectra with \ion{He}{2} Ly$\alpha$ breaks, using all ACS spectra in S09a,b and Z08 with significant flux breaks, except those with LLS (since they have extremely low S/N continua, and are heavily affected by hydrogen absorption).
We separate the spectra into four evenly-spaced redshift bins: $3.1 < z < 3.3$ (7 objects), $3.3 < z < 3.5$ (7 objects), $3.5 < z < 3.7$ (5 objects), $3.7 < z < 3.9$ (6 objects).
Not all spectra contribute to every wavelength, but all contribute at the \ion{He}{2}~Ly$\alpha$ break.
To find the error on the median, we use a bootstrap method over the medians of all contributing spectra at each wavelength, smoothed with a running average over three pixels; for further discussion of our bootstrap methods, see section \ref{sec:GP_Trough}.
The stacks are shown in figure~\ref{fig:stacked_specfits}.

\subsection{Fits of Spectra}
\label{sec:specfits}
We fit a simple model quasar spectrum with an absorption edge to the stacked spectra and a few higher S/N individual objects.
We use a power law continuum, with Gaussian emission and absorption lines, and an absorption edge.
There is not expected to be any change in spectral index over our limited wavelength region, and due to high Doppler parameters, the emission lines are approximately Gaussian rather than Voigt profiles, so this should provide a good basic fit.
We model the IGM absorption as an optical depth $\tau(z) = \tau_0/(1+w(z))$, where $\tau_0$ is the optical depth far from the quasar (where the intergalactic UV background dominates), and $w$ is the ratio of ionizing flux from the quasar and from the intergalactic background.
To determine $\tau_0$, one must have observations far from the quasar, at $w \ll 1$, which might be possible with some (though not all) of our spectra.
We take the simple model $w(z_1) = w_0/(D_L^2(z_1,z_2))$, where $D_L$ is the luminosity distance from the quasar (in practice, from the absorption edge at $z_2$).
Physically, $w_0$ depends on the luminosity and SED of the quasar, and thus is of scientific interest for future high-quality spectra.
However, the current data do not allow us to robustly determine its value, and therefore we treat it as a nuisance parameter that we allow to vary in our fits, constraining it only by a quasar lifetime limit of $10^8$ years, which should be appropriate for these luminous quasars
(\citealt{hop06}, although see \citealt{kel10}).
Simulations show that this geometric dilution factor should be a good fit to the proximity zone everywhere except $\lesssim$~1~Mpc from the quasar \citep{par10}, which is not resolvable in our spectra.
We use a $\Lambda$CDM cosmology with $h=0.704$, $\Omega_{\Lambda}=0.73$, and $\Omega_m=0.27$ \citep{kom10}, although our results are insensitive to these values, since we do treat $w_0$ as a nuisance parameter.

We fit this simple model to our spectra using the Levenberg-Marquardt least-squares fitting program MPFIT \citep{mar08}.
As a first step, we fit the observed data with just the analytic quasar spectrum described above.
We use this to find the line centers, and the wavelength of the absorption edge, and obtain initial values for line widths and amplitudes.
In the second step, we fit the data using convolution of the model through the ACS prism's (nonlinear) instrumental resolution.
Including this convolution makes the fitting step much less efficient, so we fix the wavelengths of line centers and the absorption edge at this point, as they are negligibly affected by instrumental resolution.

\begin{figure}
\epsscale{1.1}
\plotone{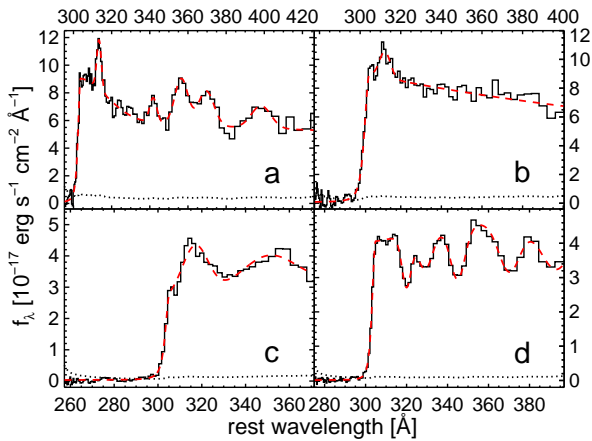}
\caption{Spectra of four individual quasars with fits overplotted (dashed lines, red in online version) and spectrum errors shown (dotted lines). (a) SDSS~J1006+3705, $z=3.20$, fit with six emission lines, including the only use of \ion{He}{2}~Ly$\alpha$ in any of our fits in this figure or figure \ref{fig:stacked_specfits}. Reduced $\chi^2=1.15$.
(b) SDSS~J1253+6817, $z=3.47$, with with one emission line. Reduced $\chi^2=0.97$.
(c) SDSS~J1711+6052, $z=3.82$, with two emission lines. Reduced $\chi^2=0.89$.
(d) SDSS~J2346-0016, $z=3.51$, with five absorption lines. Reduced $\chi^2=0.68$.}
\label{fig:individual_specfits}
\end{figure}

The fits for the S09a,b targets 1006+3705 ($z=3.2$) and 1253+6817 ($z=3.47$) are shown in figures~\ref{fig:individual_specfits}a,b.
We also include fits for the Z08 targets 1711+6052 and 2346-0016 in figures~\ref{fig:individual_specfits}c,d.
The first three fits use no absorption lines, while the 2346 fit uses \textit{only} absorption lines, with no emission lines.
Below we comment briefly on the apparent emission lines, and we consider the IGM opacities of these fits in section \ref{sec:GP_Trough}.
In every case, the break wavelength found in our fit is consistent with \ion{He}{2} at the quasar redshift.
The redshift-binned stacks are fit in figure \ref{fig:stacked_specfits}.

Due to the high opacity of the IGM  at $\lambda < 912$~\AA\ from \ion{H}{1} continuum absorption, clear sightlines to examine quasars in the extreme UV (EUV) are rare.
Our own galaxy's opacity constrains these sightlines to be at higher redshift ($z > 912$~\AA$/\lambda - 1$).
In the process of finding \ion{He}{2} quasars, we discovered a number of such clear sightlines appropriate for EUV quasar spectroscopy.
Some quasars appear to have multiple observable emission lines in this region, (see, for example, figure \ref{fig:individual_specfits}a), although it is interesting that these lines may not be as universal in line strength as the well-known lines seen in the optical and the near and far UV.

We have run simple broad emission line region models using the photoionization software Cloudy $08.00$ \citep[last described in][]{fer98}, with a standard range of density and ionization paramters \citep[e.g.,][]{kor97}, solar and supersolar metallicites, and a variety of SEDs \citep{mat87,kor97,har07}.
These models indicate that the strongest emission line in this portion of the EUV ($\lambda \sim 304$--$430$~\AA) is \ion{He}{2}~Ly$\alpha$, with an equivalent width of $\sim$$6$--$11$~\AA\ for a covering factor of 10\%.
This strength is easily resolvable in our spectra, and the line is predicted to dominate by one or two orders of magnitude over any other lines, so it is therefore puzzling that our averaged and individual quasar spectra (figures \ref{fig:stacked_specfits} and \ref{fig:individual_specfits}) do not show obvious \ion{He}{2}~Ly$\alpha$, with the exception of 1006+3705.

Our spectra may show a noticeable emission line near 313~\AA, where photoionization models predict a \ion{C}{4} line.
Although this line is predicted to be stronger than most other EUV lines, the difference is fairly small, and the models uniformly predict this line to be substantially weaker than \ion{He}{2}~Ly$\alpha$.
We find that microturbulence is an effective \citep[and plausible; e.g.,][]{bal04} way of increasing metal line strength relative to He in the EUV \citep[as has been previously noted for metals compared to H in the FUV/NUV;][]{fer99}, but it still leaves the strongest metal lines over ten times weaker than \ion{He}{2}~Ly$\alpha$.
The four well-known \ion{He}{2} quasars that have been observed at higher spectral resolution and S/N have not obviously shown any metal lines, or even \ion{He}{2}~Ly$\alpha$ consistently.
Further analysis of potential emission lines in our ACS spectra is ongoing, but it appears that definitive answers await the many \ion{He}{2} quasar spectra our team and others are obtaining with the higher resolution of COS.

\begin{figure}
\epsscale{1.0}
\plotone{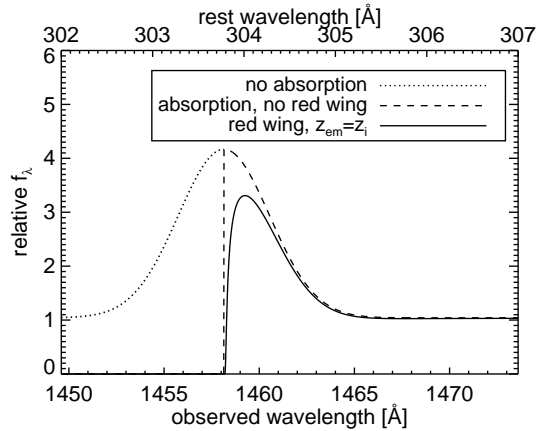}
\caption{Example of how an IGM red absorption wing affects a Gaussian \ion{He}{2}~Ly$\alpha$ emission line, for a $z=3.8$ quasar (assumed to be near the beginning of \ion{He}{2} reionization, and thus having the strongest possible red wing).
The red wing might substantially reduce the line's EW and peak flux relative to the case with absorption only below the break (the EW decreases 26\% with the inclusion of the red wing, in this example).
However, this is observationally degenerate with an intrinsically weaker line, unless the very narrow absorption profile can be seen (and distinguished from a proximity profile).
The low-resolution FUV grating of COS has a dispersion of $0.56$~\AA\ per resolution element; the highest-redshift \ion{He}{2} quasars currently known are too faint for observation with COS's medium-resolution grating.}
\label{fig:redwing_line}
\end{figure}

We do see an absorption profile reminiscent of the feature in 1711+6052 (figure \ref{fig:individual_specfits}c) in some of our individual spectra (e.g., figure \ref{fig:individual_specfits}b) and many of our averaged stacks, albeit generally not one as prominent as that of 1711.
One possible explanation is that \ion{He}{2}~Ly$\alpha$ is absorbed by the IGM, since our quasars are chosen to lie in an era with a substantial \ion{He}{2} fraction.
A neutral (or for helium, singly-ionized) IGM is expected to produce a red wing of absorption near resonant lines \citep{mir98}, as in figure \ref{fig:redwing_line}.
This is a possibility of great interest, as we now have several known high-redshift ($z>3.7$) \ion{He}{2} quasars (Z08, S09a,b), which may exist in an IGM with a substantial \ion{He}{2} fraction.
For $x_{\textrm{{\scriptsize He}}\,\textrm{{\scriptsize{\sc II}}}} \sim 1$, the Gunn-Peterson trough would be saturated \citep[although see][]{fur09}, but the red damping wing is most easily observed in just these conditions, and would be a way of directly constraining high $x_{\textrm{{\scriptsize He}}\,\textrm{{\scriptsize{\sc II}}}}$.
For our red wing predictions, we take $\Omega_b=0.0456$ \citep{kom10} and helium mass fraction $Y_p=0.2483$ \citep{ste07}.

However, as one might expect, we predict that such a red wing would be extremely difficult to detect in \ion{He}{2}.
When observing an object with an ionized proximity zone, such as many quasars, the red wing is expected to be substantially weakened and thus difficult to detect in \ion{H}{1}~Ly$\alpha$ \citep{mad00,cen00}.
For \ion{He}{2}~Ly$\alpha$, however, this red wing should be reduced even compared to \ion{H}{1}~Ly$\alpha$, 
largely because of the lower abundance of helium and the later redshift of its reionization (leading to a lower density).

A possible factor intensifying the red wing is that quasars probably form in denser regions, while analytical red-wing analyses have assumed a uniform IGM.
While moderate overdensities might still be insufficient to produce an observable red wing, substantial overdensities (such as \ion{He}{2} DLAs) might be strong enough to do so (for an example possibly involving a high-redshift galaxy, see Z08).
Denser environments would have higher absorption because of the higher density, but they also have more photon sources, and thus might be more highly ionized, reducing absorption.
Apart from the quasar proximity zone itself, however, this increase in ionization is probably not significant, because quasars are by far the dominant producers of \ion{He}{2}-ionizing photons.

While any diffuse IGM \ion{He}{2} red wing might be difficult to detect in an absorption profile (and impossible with the resolution of our present ACS spectra), it will likely reduce the flux of the \ion{He}{2}~Ly$\alpha$ line (see figure \ref{fig:redwing_line}).
Of course, this effect could not be discerned unless the intrinsic line flux were known, but it does reduce the \ion{He}{2}~Ly$\alpha$ strength relative to our photoionization models.
However, this reduction is moderate, even without an ionization zone diminishing the red wing, and could not cause the line to disappear.
The red wing effect should be larger at higher redshift, both because the IGM helium is less ionized and because the IGM density is larger.
However, we see in SDSS~J1137+6237 ($z=3.78$; S09b) what certainly appears to be a \ion{He}{2}~Ly$\alpha$ emission line that rises right until the \ion{He}{2} Gunn-Peterson break.
This confirms our calculation that IGM red wing effects, if present, are small and not resolved by the ACS/SBC prism.

It might be possible to observe this red wing absorption with COS, given the fortuitous circumstances of a relatively bright \ion{He}{2} quasar at high enough redshift to be near the onset of \ion{He}{2} reionization, and without a substantial proximity zone.
Such requirements are not impossible to meet.
For example, the very luminous quasar HE~2347-4342 shows no proximity zone \citep{shu10}, although it is at too low a redshift to be of use for this test.
The known high-redshift \ion{He}{2} quasars are quite faint, and require using COS's low-resolution grating, which has a dispersion of $\simeq 0.6$~\AA\ (barely able to distinguish the red wing from average-density IGM, in a noise-free spectrum; see figure \ref{fig:redwing_line}).
Brighter quasars, should some be discovered at $z>3.7$, or should the epoch of \ion{He}{2} reionization start later than expected, could use the medium-resolution grating of COS, with a dispersion of $0.07$~\AA.
One advantage of looking for the red damping wing during the \ion{He}{2} reionization epoch, compared to that of \ion{H}{1}, is that there is a known substantial quasar population at the presumptive beginning of helium reionization ($z \sim 4$), but not at the beginning of hydrogen reionization.

There is another IGM absorption effect reducing \ion{He}{2}~Ly$\alpha$ emission that does not require a strong red wing.
It is known that the high-ionization emission lines of quasars are systematically blueshifted relative to low-ionization lines \citep[e.g.,][]{gas82,ric02}, and thus presumably also relative to the systematic redshift of the quasar and neighboring IGM.
For example, \citet{ric02} find that \ion{C}{4} 1549 is shifted by a median value of nearly 1000~km~s$^{-1}$ relative to \ion{Mg}{2} 2799, and in certain instances the shift may be up to $\sim$3000~km~s$^{-1}$.
Shifts such as this would move \ion{He}{2}~Ly$\alpha$ emission line substantially into the region covered by IGM \ion{He}{2} absorption, without relying on the red wing for absorption.
However, the same ionized proximity zone that destroys the red wing also impacts this effect.
It should be noted that of the four well-studied \ion{He}{2} quasars, Q0302-003 and PKS1935-692 have clear proximity zones \citep{hea00,and99} while HS~1700+6416 and HE~2347-4342 do not \citep{fec06,sme02,shu10}.
Despite this difference, none of the four have clear \ion{He}{2}~Ly$\alpha$ emission lines, even in higher-resolution spectra.
HS~1700+6416, at $z=2.72$, is in an IGM where \ion{He}{2} is substantially reionized, and thus the IGM opacity is much smaller.
Even so, the quasar does not show clear \ion{He}{2}~Ly$\alpha$ emission.
Thus, while the blueshifting of emission lines might be a contributing factor to the weakness of \ion{He}{2}~Ly$\alpha$ in some objects, it seems unlikely to explain the whole effect alone.

Because of the small sample size, there is always the possibility that some of the variation we see in emission lines and absorption profiles is just due to intrinsic differences between quasars.
We do see some objects with apparent \ion{He}{2}~Ly$\alpha$ emission, sometimes strong (e.g., SDSS~J1442+0920 and SDSS~J1137+6237, both S09b, and SDSS~J1315+4856, S09a).
We have enough objects that we know strong \ion{He}{2}~Ly$\alpha$ emission is not common, as the stacks of section \ref{sec:spectrum_stacks} show, but we await higher-resolution COS spectra to distinguish the effects of emission and absorption.

\section{IGM \ion{He}{2} Opacity}
\label{sec:igm_opacity}

In this section we consider some ways of combining our spectra to extract information about the IGM \ion{He}{2} opacity.
At the most basic level, our spectral data allow for helium opacity measures along several individual sightlines, as well as an ensemble measure of the median optical depth averaged over many sightlines.
With a large number of high-redshift sightlines clean down to \ion{He}{2}, it is also natural to see if our data reveal any redshift evolution of the \ion{He}{2} Gunn-Peterson opacity.
There is some evidence from \ion{H}{1}~Ly$\alpha$ forest observations and inferred temperatures that \ion{He}{2} reionization occurred earlier than some previous studies suggested, at $z \sim 3.4$ \citep{lid09}.
Although this result is interesting, \citet{lid09} caution that it is statistically marginal, and so it would be useful to directly examine the Gunn-Peterson opacity to constrain this.
The ACS/SBC spectra taken in our reconnaissance programs were intended to verify clean sightlines only down to \ion{He}{2}~Ly$\alpha$, and as a result, some of our objects have very little data below the \ion{He}{2} break.
While detailed analysis of the Gunn-Peterson trough in those particular objects is not possible, our large sample includes many objects at higher redshift that do allow for such an analysis.

\subsection{\ion{He}{2} Gunn-Peterson Trough}
\label{sec:GP_Trough}

The first measure of opacity we turn to is that obtained in our spectrum fits of individual objects (figure \ref{fig:individual_specfits}).
No region of our spectrum for 1006+3705 ($z=3.20$) is in the $w \ll 1$ case (i.e., where the IGM ionizing background dominates over the quasar flux), and we therefore have no good determination of $\tau_0$.
There is somewhat more hope for 1253+6817 ($z=3.47$), as we have a long trough (although with low S/N).
Our fit finds $\tau_0 \sim 5$ far from this quasar, at a mean redshift of $z \simeq 3.25$, while at a similar redshift, 2346-0016 ($z=3.51$) is fit with $\tau_0 \sim 6$.
We fit 1711+6052 ($z=3.82$) with $\tau_0 \sim 4$, lower than those at lower redshift, but this difference is unlikely to be significant.
There are large systematic uncertainties in these measurements, arising both from our choice of model and from the expected substantial sightline variance \citep[e.g.,][]{fur10}, and we therefore present these values only as a description of our fits, and not as reliable measurements of the IGM opacity.

Fitting the stacks (figure \ref{fig:stacked_specfits}) provides another estimate of the IGM opacity, one that averages over sightline variance.
We formally obtain fits for the opacity that are not unreasonable ($\tau_0 \sim 4$--$6$, consistent with our more robust estimates found below), but the systematic errors associated with these determinations are still very large, and limit their usefulness.
The model we are fitting to the data is also simplistic, in that it assumes a constant opacity in the Gunn-Peterson trough of any given spectrum, with no redshift evolution.
We turn, therefore, to our entire data set as a collection of many opacity data points, allowing for redshift evolution, and averaging over individual objects, spectrum extractions, and exposures.

We pursue here an analysis of average \ion{He}{2} Gunn-Peterson opacity in the redshift range $z=3.0$--$3.7$, binned by $\Delta z=0.1$ to look for redshift evolution.
\citet{dix09} looked for evolution in the range $z=2.0$--$3.2$, using literature values for the handful of well-known sightlines with data covering some of this region (at most redshifts in this range, only one or two quasars contributed data).
While we have the advantage of a far greater number of sightlines, we also have the disadvantage of much lower-resolution, lower-S/N spectra, although the inclusion of the longer exposures of Z08 helps somewhat with the S/N.
(We note, however, that an analysis done without these two long exposures still gives results consistent with those presented here.)
The background subtraction for the ACS prism, while generally acceptable, is a source of considerable uncertainty when extracting such a small signal.
In creating their lower-redshift opacity evolution analysis, \citet{dix09} were able to ignore those literature opacity values that were only lower limits.
However, due to the relatively high noise of our short exposures, as well as the higher opacity one would expect at this higher redshift regime (prior to the completion of \ion{He}{2} reionization), we cannot ignore data points that only supply lower bounds on the opacity.

For each object, we define a portion of the spectrum to be used as its Gunn-Peterson trough, excluding the very noisy data at $\lambda < 1250$~\AA, as well as any obvious ionization zones near the quasar's \ion{He}{2}~Ly$\alpha$ break.
We fit a power law to the continuum region (avoiding the region around any \ion{He}{2}~Ly$\alpha$ emission), and extrapolate this to find the expected continuum flux in the absence of IGM helium absorption, $f_c(\lambda)$.
Each data point in the Gunn-Peterson trough has a signal flux, $f_{\textrm{{\scriptsize GP}}}$ (after the background is subtracted during spectrum extraction).
We find the ratio $f_{\textrm{{\scriptsize GP}}}(\lambda)/f_c(\lambda)$, and average all ratios in each redshift bin.
By working with the ratio, we avoid absolute flux calibration problems, which might be significant for faint objects in these prism exposures (S09a).
This approach also avoids the highly skewed distribution of the optical depth, where, because of large or effectively infinite opacities, the mean and median are very different, with the former generally not defined.
Even the ratio distribution is skewed, however, so we prefer the median as a much more robust estimator than the mean.

Because the signal ranges from being of the order of the background to much smaller, we have a number of unphysical data points for which $f_{\textrm{{\scriptsize GP}}} < 0$, simply due to fluctuations in the background.
In order to not bias the average, a standard procedure is to include these unphysical values while taking the average \citep[e.g.,][pp. 136--142]{cow98}, and we use this for our point estimates of the opacity.
However, with lower numbers of data points, and $f_{\textrm{{\scriptsize GP}}} \sim 0$, even the average can attain unphysical values.
In this case, no point estimate exists for the opacity, but we would still like to obtain one-sided confidence intervals.
The question of how to find reasonable confidence intervals for data near physical boundaries is quite difficult, even for very simple probability density functions (p.d.f.s) \citep[e.g., for a review of methods to deal with Gaussian and Poissonian data near boundaries, and their difficulties, see][]{man02}.

Since the p.d.f. of our opacity is both complicated and substantially unknown, we take a non-parametric bootstrap Monte Carlo approach. 
To ensure adequate data for resampling, we require $n \geq 20$ data points in each redshift bin (in practice we have $n \geq 24$), which means that we must discard the poorly populated $z=3.7$--$3.8$ bin.
In each bin we perform $N=10^5$ bootstrap realizations.
Because the $\log$ function is monotonic, the confidence interval found for the ratio distribution transforms with equivalent cover to an interval for opacity.
Note, however, that while this transformation invariance is true for standard bootstrap percentile confidence intervals, it is {\it not} true for percentile-$t$ methods \citep{dic96}.
Thus despite the latter technique being better in some circumstances, we avoid it here.
The statistical analysis we perform has some similarities to that independently created by \citet{pre93}, although they were not working near the zero bound and could discard negative flux measurements.

In standard bootstrapping, which is a technique of resampling with replacement, every data set generated by the bootstrap contains $n$ data points, where $n$ is the number of original data points observed.
The statistics of this ensemble of mock data sets can then be calculated, and assumed to approximate the unknown p.d.f.
A simple but quite broadly applicable method of protecting the bootstrap against failure is the $m$-out-of-$n$ method of bootstrapping, where each bootstrap data set is generated with $m<n$ data points \citep{che08}.
The required asymptotic behavior of $m(n)$ is that as $n \to \infty$, $m \to \infty$ and $m/n \to 0$ (e.g., $m=n^{\alpha}$, for $0< \alpha < 1$).
In practice, our specific choice of $m$ is found via the minimum volatility method \citep[][pp. 197--200]{pol99}.
While our data are not subject to most of the problems the $m$-out-of-$n$ method addresses, there is at least one case in which it might be helpful.
For a signal at zero, one expects about half of the data points to lie in the unphysical negative regime (after background subtraction).
Although the median is a fairly robust estimator, it has a breakdown point of 50\%, and so when it happens that more than half the points in a bin are unphysical, the median itself becomes unphysical.
This problem occurs in our $z=3.1$--$3.2$ bin, where $\sim$53\% of the data are unphysical---not enough to think that this is due to anything other than a random fluctuation (18\% probability) around a very small (consistent with zero) signal, but enough to interfere with the determination of our confidence intervals.
In this case, confidence intervals with larger coverage help, but so does the $m$-out-of-$n$ approach.
We conservatively display $m$-out-of-$n$ confidence intervals for all redshift bins, although the difference from the $n$-out-of-$n$ intervals is insignificant in most cases. 

\begin{figure}
\epsscale{1.1}
\plotone{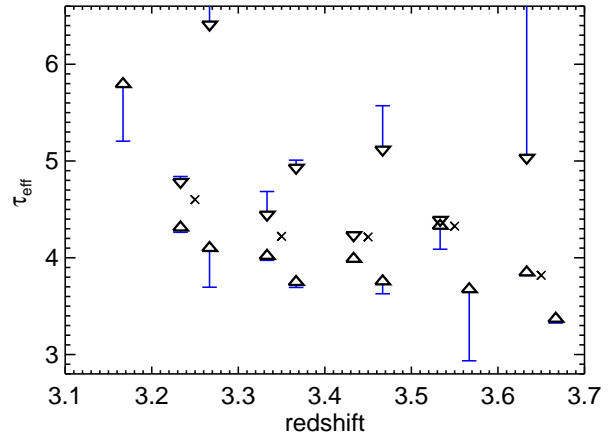}
\caption{\ion{He}{2} Gunn-Peterson opacity as a function of redshift, as derived from our ACS spectra.
Where available, median estimates are plotted as `x's, and confidence interval limits are plotted slightly offset in the horizontal direction.
In each bin, where available, offset to the left are the 68\% confidence limits, and offset to the right are the 95\% limits.
Upper and lower limits are calculated independently; note that when combining independent $1-\alpha$ upper and lower limits, the resulting confidence interval is $1-2\alpha$.
The thick arrowheads denote standard bootstrapping limits, while the thinner blue tails are for $m$-out-of-$n$ methods (see text).
The apparently anomalous opacity of the lowest redshift bin is not a statistically significant deviation from the other bins (see text for a discussion).}
\label{fig:gp_opacity}
\end{figure}

Figure \ref{fig:gp_opacity} shows Gunn-Peterson opacity versus redshift, and it is clear that no statistically significant evolution can be seen.
Where available, the median estimate is plotted as an `x', and the confidence limits as arrows (offset slightly in the horizontal direction for clarity).
The thicker black arrows indicate standard bootstrap lower limits, 68\% for those on the left and 95\% for those on the right, while the thinner blue lines show 68\% and 95\% confidence limits for $m$-out-of-$n$ bootstrap estimates.
To enable comparison between redshift bins, only one-sided $1-\alpha$ limits of the opacity are plotted, so in the cases where finite two-sided confidence intervals exist, when one combines the upper and lower limits shown it produces a $1-2 \alpha$ two-sided confidence interval.
While the $z=3.1$--$3.2$ bin looks anomalous, this appears to be due to the data problems discussed above.
The $z=3.2$--$3.3$ bin has upper limits defined for most confidence levels, and we can therefore state that when considering the two-sided 90\% intervals, there is no significant difference between this bin and that of $z=3.1$--$3.2$; i.e., the apparent anomaly is not statistically signficant.
We conclude that our data are unable to reveal any evidence of opacity evolution in this redshift regime (a conclusion confirmed by shifting bin edges and sizes), although we caution that this says more about our data than it does about the IGM---note that three of the six bins (including those at highest redshift) include infinite opacity in their two-sided 90\% confidence intervals.

We aggregate the data further to make an opacity estimate for this entire high redshift region, which is mostly inaccessible to the well-known \ion{He}{2} quasar sightlines.
In the large bin $z=3.1$--$3.7$, we find a median $\tau=4.90^{+1.86}_{-0.64}$ (95\% confidence, $m$-out-of-$n$).
Because the ACS prism provides more data points at lower wavelength (and thus redshift), the mean redshift of the data used for this opacity is $z \approx 3.3$.
This opacity at this redshift agrees well with the $\tau=4.75^{+0.41}_{-0.29}$ measurement of Q0302--003, found for $z \approx 3.1$--$3.2$ \citep{hea00}, and the weak lower limits of $\tau=6^{+\infty}_{-3}$ for $z \approx 3.4$--$3.5$ \citep{zhe04a} and $\tau > 3.1$ (90\% confidence) for $z \approx 3.35$ (Z08).
Comparing our binned data of figure \ref{fig:gp_opacity} to the models considered in figure 7 of \citet{dix09}, we find that our measured opacity is consistent with their model of \ion{He}{2} reionization ending (full ionization) at $z_{{\rm rei}} \sim 3$, and {\it not} consistent with the extreme model of $z_{{\rm rei}}>3.8$ (it is also inconsistent with an extrapolation of the other extreme model, $z_{{\rm rei}}=2.5$).
We caution, however, that the models of \citet{dix09} require assumptions about the spectral index of quasars in the EUV.
This is poorly known at $\lambda < 912$~\AA\ \citep{tel02,sco04}, and has no direct observational constraint at the energies needed to ionize \ion{He}{2}, $\lambda < 228$~\AA.
This uncertainty relaxes the constraints we can place on the redshift of full helium reionization \citep{mei05}.

The ACS spectra here, with the exception of those from Z08, were obtained with the intent of verifying the existence of \ion{He}{2} Gunn-Peterson troughs.
As such, it was not expected that they themselves would be a particularly sensitive probe of the \ion{He}{2} opacity, but it was worthwhile investigating this question, as they constitute by far the largest sample of \ion{He}{2} quasars to date.
Indeed, beyond $z \sim 3.2$ the only opacity data is from these ACS spectra (including the Z08 objects in our sample) and the weak limit of \citet{zhe04a} (which was from SDSS~J2346-0016, reobserved and included at higher S/N in our sample).
It is therefore perhaps disappointing, but hardly surprising, that the uncertainties in the current data do not allow any robust determination of the \ion{He}{2} opacity evolution in this interesting redshift regime.
Higher S/N spectra of the brighter of our quasars will be invaluable for making this direct determination of opacity; a number of such observations are now planned for \textit{HST}.

\section{Conclusion}
\label{sec:conclusion}

The relatively large numbers of \ion{He}{2} quasars our team has recently found allow us to construct average spectra in four high-redshift bins.
While the individual objects are often low S/N, one of the primary sources of uncertainty in \ion{He}{2} Gunn-Peterson studies is sightline variance, which we can now begin to average over.
We find that the absorption profile of SDSS~J1711+6052 may not be as unusual as was initially thought.
We cannot resolve the puzzle of why \ion{He}{2}~Ly$\alpha$ emission appears so weak in many (though not all) known \ion{He}{2} quasars, but we consider two possibilities, and find that neither a red wing of IGM absorption nor emission-line blueshifting can alone account for this.
It is unlikely to find \ion{He}{2} quasars that will allow observations of the diffuse IGM red wing at the beginning of \ion{He}{2} reionization, but denser IGM knots and rare young quasars might provide such opportunities.

We find an IGM \ion{He}{2} $\tau_{\textrm{{\scriptsize GP}}} \sim 5$ at $z=3.1$--$3.7$, in broad agreement with the few other probes of this high-redshift regime.
Our data are not of sufficient quality to constrain the redshift evolution of this opacity to any significant degree, although we may be able to rule out models in which \ion{He}{2} reionization ends extremely early ($z \gtrsim 3.8$) or extremely late ($z \sim 2.5$), assuming particular quasar SEDs.
Estimates of the Gunn-Peterson opacity derived from model fits of single, higher-S/N quasars agree with those found from ensemble statistics.

The discovery of more \ion{He}{2} quasars will allow further reduction in sample variance, and higher-resolution, higher-S/N observations with COS will allow better constraint on both absorption profiles and \ion{He}{2} opacities.
Observations with both these aims are currently underway.

\acknowledgments

We gratefully acknowledge support from NASA/{\it GALEX} Guest Investigator grants NNG06GD03G and NNX09AF91G.
Support for {\it HST} Program numbers 10907, 11215, and 11982 was provided by NASA through grants from the Space Telescope Science Institute, which is operated by the Association of Universities for Research in Astronomy, Incorporated, under NASA contract NAS5-26555.

The authors thank Gary Ferland and Robin Evans for their helpful comments on aspects of this paper.

\clearpage

\end{document}